\documentclass[pre,showpacs,amsmath,amssymb,longbibliography]{revtex4-2}
\usepackage{amsmath,amssymb,latexsym,epsfig,graphics,epsf, import}
\usepackage{hyperref}
\usepackage{subcaption}
\usepackage{graphics}
\usepackage{float}
\usepackage{footnote}
\usepackage[dvipsnames]{xcolor}
\usepackage{graphicx,float, hyperref, cleveref, import, subcaption, setspace}
\DeclareUnicodeCharacter{2009}{\,}
\newcommand{\ea}[1]{\left\langle#1\right\rangle} 
\newcommand{\eq}[1]{Eq.~(\ref{#1})} 
\newcommand{\Eq}[1]{Equation~(\ref{#1})}
\newcommand{\be}{\begin{equation}}
\newcommand{\ee}{\end{equation}}
\newcommand{\Fig}[1]{Figure~\ref{#1}}
\newcommand{\fig}[1]{Fig.~\ref{#1}}

\newcommand{\mycomment}[1]{}
\begin{document}

    \title{Large temperature-up-jump simulations of a binary Lennard-Jones system}
    \date{\today}
    \author{Aude Y. Amari}\email{aude.amari@gmail.com}
    \affiliation{\textit{Glass and Time}, IMFUFA, Department of Science and Environment, Roskilde University, P.O. Box 260, DK-4000 Roskilde, Denmark}
    \author{Lorenzo Costigliola}\email{lorenzo.costigliola@gmail.com}
    \affiliation{\textit{Glass and Time}, IMFUFA, Department of Science and Environment, Roskilde University, P.O. Box 260, DK-4000 Roskilde, Denmark}
    \author{Jeppe C. Dyre}\email{dyre@ruc.dk}
    \affiliation{\textit{Glass and Time}, IMFUFA, Department of Science and Environment, Roskilde University, P.O. Box 260, DK-4000 Roskilde, Denmark}

\begin{abstract}
This paper presents simulations of the physical aging of a binary Kob-Andersen-type Lennard-Jones liquid following large temperature up-jumps from equilibrated states of high relaxation time. The purpose is to investigate how well the Tool-Narayanaswamy (TN) material-time concept works for this rather extreme case of aging. First the triangular relation of the potential energy is investigated. This is found to be well obeyed, making it possible to define a potential-energy-based material time $\xi$. We proceed to study aging toward equilibrium at the final temperature 0.48 for jumps from the two temperatures 0.43 and 0.37 (primarily), monitoring the following five quantities: the potential energy, the self-intermediate scattering function, the mean-square displacement, the dynamic susceptibility $\chi_4$, and the non-Gaussian parameter $\alpha_2$. The TN material-time prediction is that all time-autocorrelation functions should collapse to only depend on the material-time difference $\xi_2-\xi_1$. This is found to work better for the $0.43\to 0.48$ temperature jump than for the $0.37\to 0.48$ jump. Our findings thus confirm the general understanding that the TN aging formalism works best for systems that are never very far from equilibrium. This raises two questions for future work: Is the collapse significantly improved if each aging quantity is allowed its own material time? Can better collapse be obtained if the material-time is generalized to be locally defined (in order to reflect dynamic heterogeneity)?
\end{abstract}

\maketitle
\vspace{1cm}

\section{Introduction}

It has been known for many years that physical aging of glasses is generally well described by the so-called material time \cite{maz77,scherer,hec10,hec15,meh21,rie22,moc24}. This concept was introduced in 1971 by Ford Motor Co. engineer Narayanaswamy for optimizing the production of windscreens \cite{nar71}. It is now recognized that the material-time approach rationalizes aging data of non-crystalline materials as different as oxide glasses \cite{scherer}, chalcogenide glasses \cite{mic16}, polymers \cite{hod95,hut95,can13,mck17}, and metallic glasses \cite{rut17,son20}. The resulting so-called Tool-Narayanaswamy (TN) or TNM (Tool-Narayanaswamy-Moynihan) formalism has been used routinely for many years in industry in connection with glass production, as well as for predicting the long-time aging taking place in the use of glasses \cite{scherer,mauro}.

\begin{figure}[H]
    \centering
    \includegraphics[width=0.9\linewidth]{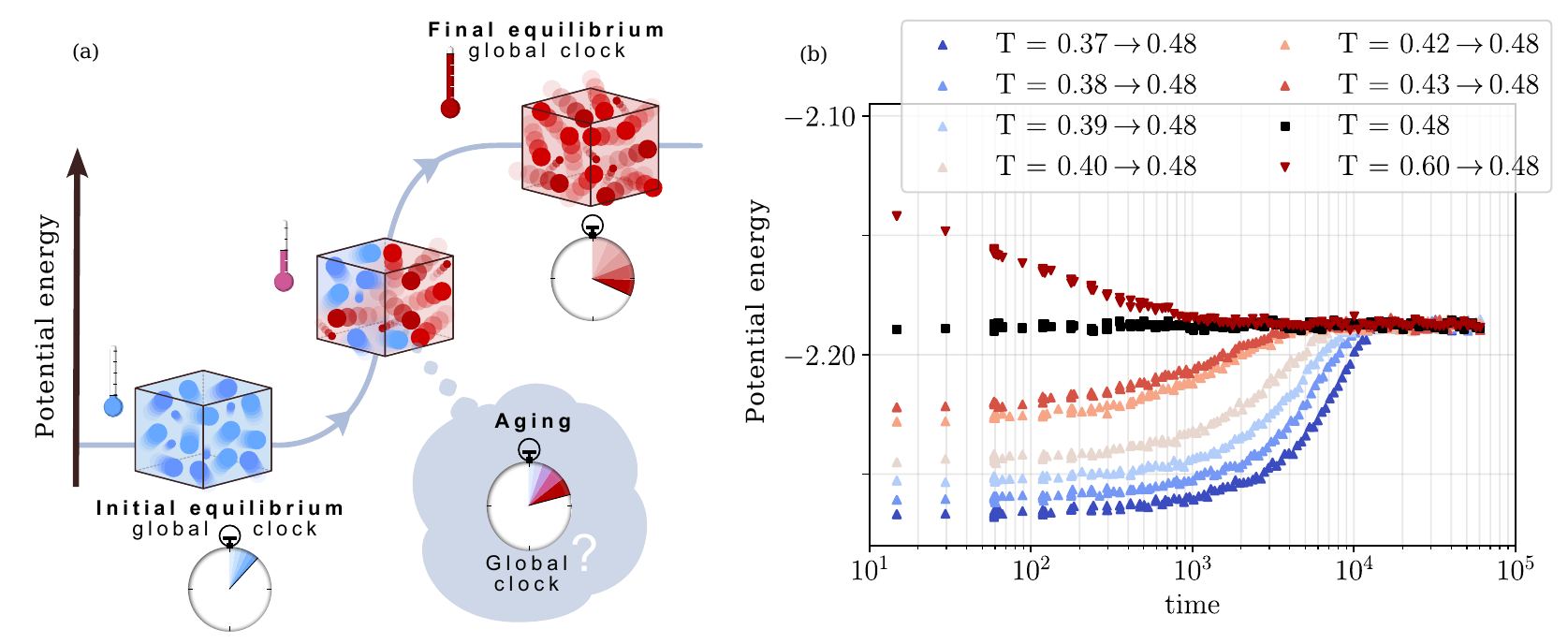}
    \caption{Physical aging. 
    (a) Schematic drawing of a temperature up jump starting and ending in thermal equilibrium (with time on the implicit x-axis). Even though not all quantities relax in identical manner, the standard TN aging formalism operates with a single ``global'' clock controlling all relaxations toward equilibrium.
    (b) Results for the potential energy per particle following one down jump and several up jumps to the temperature $0.48$, illustrating the noted ``asymmetry of approach'' observed in all aging simulations and experiments. The black filled rectangles mark the potential energy for an equilibrium simulation at $T=0.48$.}    \label{fig:visual_abstract}
\end{figure}

Both production and subsequent applications of glasses involve complex temperature protocols. In contrast, the conceptually simplest cases of physical aging are realized after rapid jumps from a state of equilibrium at one temperature to a final temperature that is subsequently kept constant for a time long enough to ensure equilibration \cite{rie22,vil23, hen24,lan26}. Ideally, the ``annealing'' temperature is stabilized throughout the sample before any significant relaxation has taken place \cite{hec10}. Following a temperature jump, one or more properties of the system are monitored continuously as it equilibrates, compare \fig{fig:visual_abstract}(a) that illustrates schematically aging following a temperature up jump. Due to the slow rate of heat conduction, such ideal equilibrium-to-equilibrium jumps are difficult to realize in experiments (though not impossible \cite{hec10,rie22,hen24}). In simulations, on the other hand, temperature jumps are easily implemented -- here the challenge is the different one of doing the very long simulations necessary to probe the relaxing region. Furthermore, if a global quantity is monitored, i.e., data summarized into a single number, it is difficult to obtain good statistics since the standard procedure of improving statistics by time averaging cannot be used due to the breaking of time-translational invariance.

In both simulations and experiments it is universally observed that temperature up and down jumps result in quite different responses. A  temperature-down jump equilibrates faster and is more ``stretched'' in time than an up jump to the same final temperature; the latter is ``compressed'' in time when it eventually picks up speed. This difference is referred to as ``asymmetry of approach'' \cite{kov63,mck17,mal24,nis20}, with down and up jumps termed ``self-retarded'' and ``self-catalyzed'', respectively \cite{can13}. \Fig{fig:visual_abstract}(b) gives examples of data for the potential energy following one down jump and several up jumps to the same final temperature. The relaxation time changes by a factor of 10 and $10^4$ for the smallest and largest jumps, respectively. The asymmetry between up and down jumps means that aging -- from a mathematical point of view -- is highly nonlinear. This reflects the long known fact \cite{too46} that during a temperature down jump, the structure gradually involves states of lower and lower energy, resulting in the dynamics slowing down, while the opposite happens for an up jump. 

The TN formalism goes significantly beyond this qualitative reasoning known for almost a century \cite{scherer,mck17} by predicting that a \textit{material time} $\xi=\xi(t)$ exists such that all temperature jumps to the same temperature are given by the same function of $\xi$ (for the quantity monitored). This implies that if (laboratory) time is replaced by material time, physical aging is described as a standard linear-response convolution integral over the temperature-change history \cite{nar71,scherer,dou22}. The conversion of the highly nonlinear aging phenomenon into a linear one constituted Narayanaswamy's 1971 conceptual breakthrough \cite{nar71}. 

This paper presents a numerical investigation of the Kob-Andersen binary Lennard-Jones (LJ) mixture (slightly modified to avoid crystallization), following large temperature up jumps from states of equilibrium. The purpose is to investigate the limits of the TN material-time description for such jumps. We define below the material time $\xi$ from the potential-energy time-autocorrelation function and investigate to which degree each of five different time-correlation functions collapse to be unique functions of the material-time difference $\xi_2-\xi_1$ during the aging. We first discuss briefly the system simulated, as well as the triangular relation of Cugliandolo and Kurchan \cite{cug94} and its connection to the material time: A material time only exists if the triangular relation applies. This implication was recently studied in a paper of ours introducing a ``distance-as-time'' approach to aging, in which the triangular relation was investigated in relation to different distance measures in configuration space \cite{dou22}. We here take the alternative approach of validating the triangular relation directly from potential-energy data.

\section{System simulated}

We simulated a binary Lennard-Jones (mBLJ) model, which  has the same parameters as -- and physics very similar to -- the standard Kob-Andersen model \cite{kob95} but is at least 100 times less prone to crystallization \cite{sch20}. The model is a mixture of 80\% large particles (A) and 20\% small particles (B) interacting by Lennard-Jones pair potentials, $v(r)=4\varepsilon((r/\sigma_{\alpha\beta})^{-12}-(r/\sigma_{\alpha\beta})^{-6})$ ($\alpha,\beta\in\{A,B\}$), in which $\sigma_{AA}=1$, $\sigma_{AB}=0.8$, $\sigma_{BB}=0.88$, $\varepsilon_{AA}=1$, $\varepsilon_{AB}=1.5$, $\varepsilon_{BB}=0.5$. The difference to the standard Kob-Andersen model is only that instead of using shifted-potential cutoffs at relatively long distances, a shifted-force cutoff at 1.5 is used for the AA and BB interactions. This considerably weakens the like-particle attractions, which impedes crystallization taking place by phase separation into a pure A phase \cite{ped18}. The AB interaction also has a shifted-force cutoff, but at 2.5 \cite{kob95}. All simulations were carried out with $N_A=6400$ A particles and $N_B=1600$ B particles. The simulations all lasted for $2^{23}$ time steps of size $dt=0.005$. The quantities evaluated refer exclusively to the A particles because these dominate the structural relaxation. The simulations utilized the GPU-optimized MD code RUMD \cite{RUMD}.

\section{Triangular relation}

In theoretical developments more than two decades after Narayanaswamy, which took place independent of the experimentally motivated TN formalism, Cugliandolo and Kurchan (CK) devised a general theory of aging. Central in this  is the concept of \textit{time-reparametrization invariance} \cite{cug94}. The object of study was a spin glass quenched to zero temperature from a high-temperature equilibrium state. Such a system ages continuously and, in fact, never reaches thermal equilibrium. This is in stark contrast to the TN formalism that focuses on relatively small deviations from equilibrium \cite{cug97}. Nevertheless, the CK aging theory is consistent with the existence of a material time (see below). The CK theory is based on the \textit{triangular relation} \cite{cug94}, which provides the precise definition of time-reparametrization invariance. 

The time-autocorrelation function $C_{12}$ of a quantity $A(t)$ is defined by 

\be\label{eq:C_def}
C_{12}\equiv\langle A(t_1)A(t_2)\rangle-\langle A(t_1)\rangle\langle A(t_2)\rangle\,.
\ee
Here the angular brackets denote an ensemble average (\textit{not} a time average because time-translational invariance does not apply during aging). We assume that all time-autocorrelation functions are positive and go to zero monotonically at large times. The triangular relation states that for all times $t_1<t_2<t_3$ during the aging, any two time-autocorrelation functions determine the third. Specifically,

\be\label{eq:triang}
C_{13}=F_A(C_{12},C_{23})
\ee
in which $F_A(x,y)$ is a function that in general depends on $A$. Note that \eq{eq:triang} applies rigorously in thermal equilibrium by time-translational invariance: $C_{12}$ is a function of $t_2-t_1$ and $C_{23}$ is a function of $t_3-t_2$, which implies that $C_{12}$ and $C_{23}$ determine $t_3-t_1=t_3-t_2+t_2-t_1$ that determines $C_{13}$. CK derived \eq{eq:triang} from a mean-field equation for the time development of the time-autocorrelation function. In that equation, if there is a clear time-scale separation between the fast and slow degrees of freedom, then \eq{eq:triang} applies to a good approximation at long times, i.e., for the slow degrees of freedom. 

The material time $\xi(t)$ is defined from 

\be\label{eq:xi_def}
C_{12}=\phi_{\rm eq}(\xi(t_2)-\xi(t_1))\,
\ee
in which the function $\phi_{\rm eq}$ gives the thermal-equilibrium time-autocorrelation functions via $C_{12}=\phi_{\rm eq}(t_2-t_1)$ at some reference temperature (changing the reference temperature just scales $\xi(t)$ by a constant). \Eq{eq:triang} ensures consistency of this definition by reasoning as above \cite{dou22}: $\xi(t_2)-\xi(t_1)$ determines $C_{12}$ and $\xi(t_3)-\xi(t_2)$ determines $C_{23}$, meaning that $\xi(t_3)-\xi(t_1)=\xi(t_3)-\xi(t_2)+\xi(t_2)-\xi(t_1)$ determines $C_{13}$. Each quantity $A$ can have its own material time. In this paper, however, we define the material time from the potential energy time-autocorrelation function and reparametrize all two-point functions in terms of it. 

As mentioned, the main characteristic of the TN formalism is that when time is replaced by material time, the highly nonlinear aging phenomenon is reduced to one described by linear-response theory \cite{nar71,scherer,dou22}. The CK approach is conceptually quite different. \Eq{eq:triang} is independent of how time is measured, allowing for using any smooth monotonic function of time. This ``time-reparametrization invariance'' is central in the CK theory of aging, but runs contrary to the TN philosophy in which the material time plays a special role, namely that of reducing the problem to an extension of standard linear-response theory.

Though the definition of the material time $\xi(t)$ in \eq{eq:xi_def} is straightforward, it only recently became possible to determine this quantity experimentally \cite{boh24}; this was done from dynamic light scattering data. This has not been done before because directly measuring a time-autocorrelation function is challenging. Moreover, since the statistics cannot be improved by time averaging, one needs to do proper ensemble averaging during the aging. This was implemented in Ref. \onlinecite{boh24} by analyzing data from the millions of pixels of a camera taking 10-20 pictures per second.

\section{Validating the triangular relation and determining the material time}

We first validate the triangular relation \eq{eq:triang} in which the observed quantity is the potential energy. In order to improve the statistics we consider the potential energies $u_i$ of single A particles (defined by allocating half of each interaction energy to each of the two particles involved). The normalized potential-energy time-autocorrelation function is defined by

\be\label{eq:Cuu}
C_{uu}(t_1,t_2) \equiv \frac{1}{N_A\,\sqrt{ \nu(t_1) \, \nu(t_2)}} 
\left\langle\sum_{i=1}^{N_A} \Delta u_i(t_1) \Delta u_i(t_2)\right\rangle\,.
\ee
Here $\Delta u_i(t)$ is the potential energy of an A particle minus this quantity's ensemble average at time $t$ (that is independent of $i$), $\nu(t) \equiv \left\langle\sum_{i=1}^{N_A} (\Delta u_i(t))^2\right\rangle/N_A$, and the angular brackets denote an average over 50 simulations. We henceforth leave out the word ``normalized'' when speaking about time-autocorrelation functions and refer to $C_{uu}(t_1,t_2)$ simply as the potential-energy time-autocorrelation function, for brevity denoted as $C_{12}$.

\begin{figure}[H]
        \includegraphics[width=0.9\linewidth]{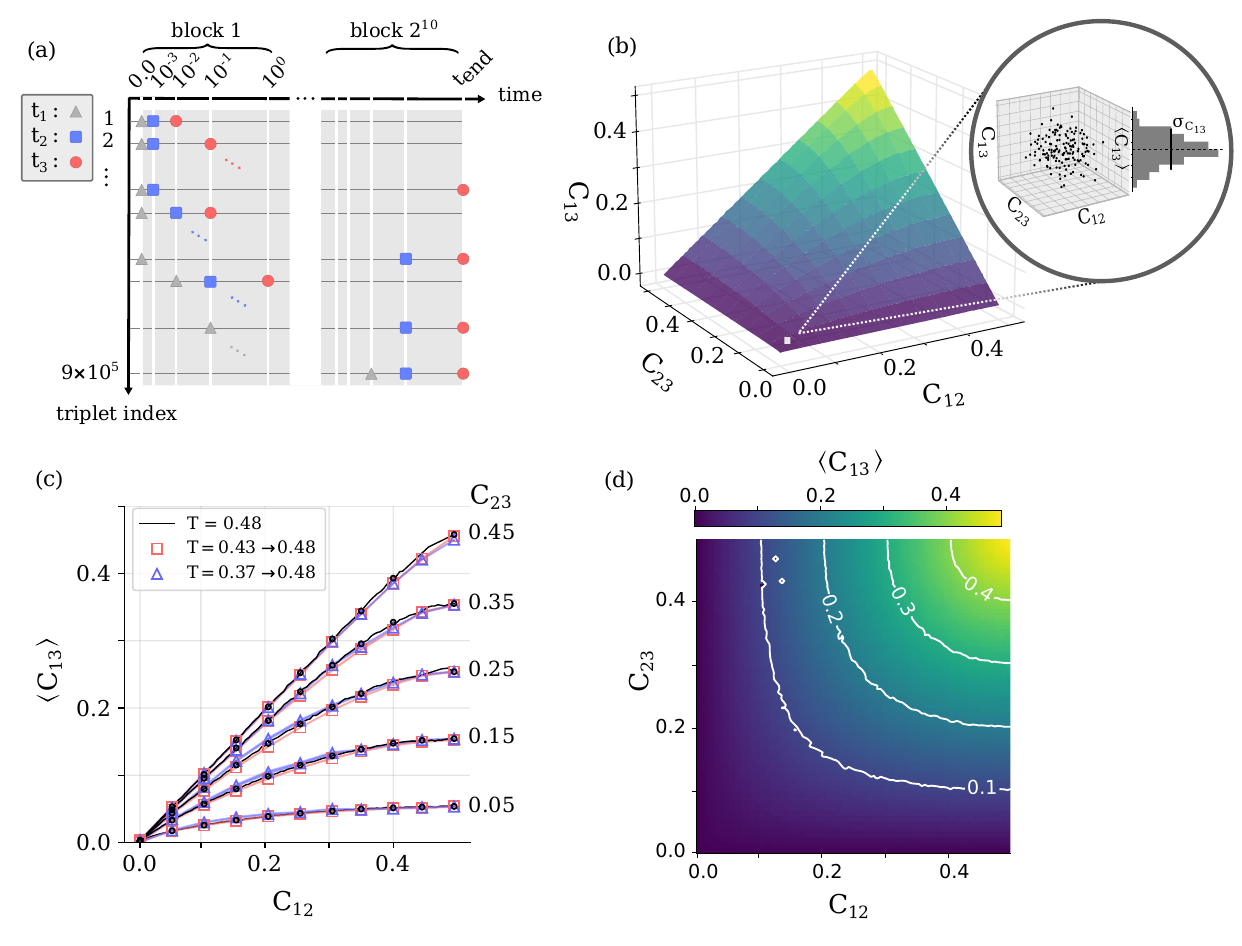}
        \caption{Parametric representation of the triangular relation, \eq{eq:triang}.
        (a) Illustration of the sampling of time triplets used to calculate correlation ``triangles''. The values on the time axis illustrate the logarithmic increase in time intervals within a single simulation block.
        (b) Illustration of the triangle binning algorithm. Each pixel contains a distribution of triangles $(C_{12}, C_{23}, C_{13})$.
        (c)  The average value $\ea{C_{13}}$ as a function of $C_{12}$ for given values of $C_{23}$. This is the mean value used to determine the function $F_A$ in \eq{eq:triang}.  The form of $F_A$ averaged over each pixel remains unchanged as the amplitude of the jump changes, as predicted by the TN formalism in which $F_A$ is determined by the equilibrium dynamics. This panel demonstrates that $F_A$ is well defined. The standard error $\sigma/\sqrt{n}$ is indicated by the width of the colored line and is smaller than the marker size ($n$ is the number of triangles and $\sigma$ the absolute standard deviation of $C_{13}$ in the pixel).
        (d) shows a heat map of the average $\ea{C_{13}}$ as a function of $C_{12}$ and $C_{23}$. }
        \label{fig:parametric_F}   
\end{figure}

The triangular relation is investigated by considering millions of time-autocorrelation functions at times $t_1<t_2<t_3$, evaluated after a temperature up jump initiated at $t=0$. For given values of $C_{12}$ and $C_{23}$ during the aging, \eq{eq:triang} posits that $C_{13}$ is uniquely determined. 250 linearly spaced values of $t_{1}$ and 60 logarithmically spaced values of $\tau$ were chosen. Denoting the set of $\tau$ values by $M$, for each value of $t_{1}$ we studied the values $t_{2} = t_{1} + \tau'$ for $\tau'\in M$, and for each $t_{2}$ we studied $t_{3} = t_{2} + \tau''$ with $\tau''\in M$ (\fig{fig:parametric_F}(a)). Of the $250*60^{2}=900,000$ $(t_1,t_2,t_3)$ triplets thus obtained we kept those for which $t_{3}$ is smaller than the duration of the simulation. 

The aging part of the relaxation response occurs at long times. Short times correspond to vibrational relaxation of the system within the cage formed by neighboring particles; here the fluctuation-dissipation theorem applies corresponding to the bath temperature $T$ \cite{cug94, cha07}. We do not show data for this regime since it is of little relevance to the present study. The reader should note, however, that reparametrization by the material time is not expected to collapse the early response (indeed it does not). 

To calculate the parametric representation of the triangular relation \eq{eq:triang}, $C_{12}$ and $C_{23}$ were each binned into 100 equal sized bins of size 0.05 units of correlation. In this way we obtain a 10,000 pixel grid, and each duplet $\left( C_{12}, C_{23} \right)$ falls into a single pixel. As shown in \fig{fig:parametric_F}(b), each pixel contains a distribution of ``triangles'' with lengths $\left(C_{12}, C_{23}, C_{13} \right)$. After disregarding pixels with less then 15 triangles, the mean value and standard deviation of $C_{13}$ at each pixel is obtained; this mean is shown in \fig{fig:parametric_F}(c)-(d). The statistics of this plot is highly sensitive to the sampling time, as well as to the change of the correlation function with $t_1$ and $t_2$. Ideally, the time sampling should be finest when the autocorrelation changes most such that a dense spectrum of values of the autocorrelation is available. We reconcile this constraint with the finite storage space available by using a saving scheme composed of many linearly spaced time blocks inside of which saving occurs logarithmically. 

\Fig{fig:parametric_F}(c) plots $C_{13}$ as a function of $C_{12}$ for given values of $C_{23}$. We learn from this plot that the shape of the function $F_A$ is virtually the same in equilibrium and for the two jumps, as predicted by the triangular relation. Physically this reflects the fact that, despite its strongly nonlinear nature, aging in some sense takes place close to equilibrium \cite{dou22}.

\begin{figure}[H]
    \centering
    \includegraphics[width=0.9\linewidth]{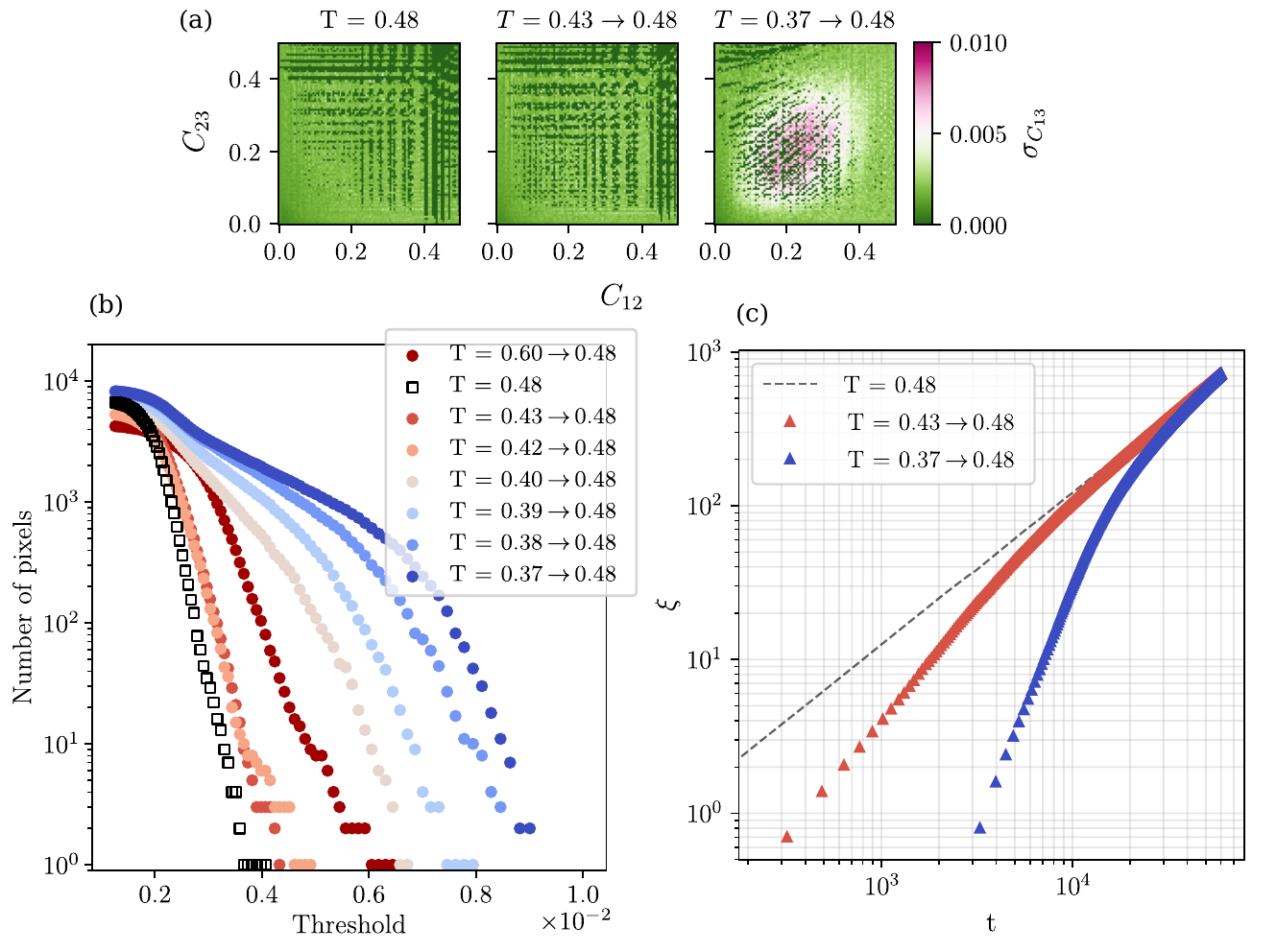}
    \caption{Numerical test of the triangular relation \eq{eq:triang}. $\sigma$ is the standard deviation of the single-particle potential energy's normalized time-autocorrelation function $C_{13}\equiv C_{uu}(t_1,t_3)$ for fixed values of $C_{12}$ and $C_{23}$. 
    (a) The standard deviation $\sigma$ is zero if \eq{eq:triang} applies rigorously. The statistical noise is illustrated by the equilibrium heat map at $T=0.48$ (left panel). In the middle and right heat maps, $\sigma$ is shown for two temperature up jumps. A small increase in $\sigma$ is observed for the large up jump.
    (b) Number of pixels whose deviation from the triangular relation exceeds a given threshold, plotted as a function of the threshold for several different temperature jumps. Although a consistent increase of the deviation is observed as the size of the jump is increased, the largest values of $\sigma$ are only roughly twice as large as in equilibrium. The initial number of pixels varies for different jumps because we disregard pixels with too little data.
    (c) The material time $\xi$ is a non-linear function of time during aging, as shown here for the two temperature up jumps focused on in Sec. V. At long times equilibrium at $T=0.48$ is approached and $\xi$ becomes proportional to time.}
    \label{fig:sigmaTR}
\end{figure}

\Fig{fig:sigmaTR}(a) summarizes our findings in the form of heat maps for equilibrium as well as the considered two temperature up jumps to $T=0.48$. The colors indicate the standard deviation defined by $\sigma^2 \equiv {\langle C_{13}^2\rangle - \langle C_{13}\rangle^2}$, used here as a measure of how well the triangular relation holds. The left panel shows results for equilibrium simulations in which the triangular relation applies rigorously as a consequences of time-translational invariance. The middle and right panels show  results for the two temperature up jumps. One of these is from the very low temperature 0.37 at which months of GPU simulations must be carried out to attain proper equilibrium configurations \cite{sch20}. Although there is a slight coloring in the latter case, the fluctuations of $C_{13}$ are minute. The deviations occur in the middle of the map corresponding to intermediate values of $t_2 - t_1$.

For several temperature up jumps and one down jump, \fig{fig:sigmaTR}(b) quantifies how far the triangular relation is from being rigorously obeyed. This is done by reporting the number of pixels deviating more than a given threshold. In general the deviations are small, albeit larger than in equilibrium at $T=0.48$ (open black squares). Since the triangular relation always applies in equilibrium, the latter data provide a baseline showing the numerical uncertainties. 

We conclude that the triangular relation is obeyed to a good approximation for the potential-energy time-autocorrelation functions. As detailed in Sec. III, this makes it possible to define a material time. This quantity we determine from the data as follows. A value of $a$ is chosen such that $\xi_2-\xi_1=1$ when $C_{12}=a$. Since the temperature jump is initiated $t=0$, we put $\xi(t=0)=0$. Next the time $t_a$ satisfying $C(0,t_a)=a$ is found, which is the time it took for the first unit of material time to elapse. Next, the time $t_{2a}$ satisfying $C(t_a, t_{2a})=a$ is found, corresponding to the time elapsed during the second unit of material time. In this way the function $\xi(t)$ is constructed iteratively from the data. We used $a=0.4$, though $a$ can of course be chosen arbitrarily (as long as it corresponds to times far from the short-time plateau). \Fig{fig:sigmaTR}(c) shows the material time as a function of time for two temperature up jumps to $T=0.48$, the two jumps that are henceforth in focus. At long times equilibrium is approached and the material time becomes proportional to time (the proportionality constant is irrelevant; it depends on the choice of $a$). At short times the material time develops more slowly with time, an effect that is of course most significant for the large up jump (blue triangles).

\section{Aging following jumps from T=0.43 and T=0.37 to T=0.48}

How well does the TN formalism work? Having established that the triangular relation to a good approximation is obeyed for the potential energy, which allows for defining a material time based on this quantity, we proceed to present results for aging of the potential energy as well as four other quantities following temperature up jumps from equilibrated states. 

The overall structural relaxation of the system is controlled by the 80\% majority particles, the A particles. The B particles are smaller and diffuse faster than the A particles for which reason we focus entirely on properties relating to the A particles. We studied the aging of the following five quantities, all referring to the A particles:

\begin{enumerate}
    \item The (normalized) potential-energy time-autocorrelation function $C_{uu}(t_1,t_2)$ defined in \eq{eq:Cuu}.        
    \item The incoherent intermediate scattering function $F_s(t_1,t_2)$ calculated from the particle displacement from time $t_1$ to time $t_2$, evaluated at the wavevector corresponding to the first maximum of the radial distribution function at the final temperature $0.48$ (the peak position was virtually constant throughout all simulations),    
    \be
    F_s(t_1, t_2) = \Bigg\langle \frac{1}{N_A} \sum_{i=1}^{N_A} \cos({\bf q}\cdot({\bf r}_i(t_2)-{\bf r}_i(t_1)) \Bigg\rangle\,.
    \ee
    \item The mean-square displacement between times $t_1$ and $t_2$,
    \be
    \ea{\Delta r^{2}}(t_1, t_2) = \Bigg\langle \frac{1}{N_A}\sum_{i=1}^{N_A}({\bf r}_i(t_2) - {\bf r}_i(t_1))^{2} \Bigg\rangle\,.
    \ee
    \item The dynamic susceptibility $\chi_4(t_1,t_2)$ quantifying the degree of dynamic heterogeneity in terms of the variance of the incoherent intermediate scattering function (even though this is a single-particle function, this quantity still reflects the volume of dynamically correlated regions and thus dynamical heterogeneity) \cite{ton05}, 
    \be\label{eq:def_chi4}
    \chi_4(t_1, t_2, q) = N_A \Biggl[ \Bigg\langle \left( \frac{1}{N_A} \sum_{i=1}^{N_A} 
    \cos({\bf q}\cdot({\bf r}_i(t_2)-{\bf r}_i(t_1)\right)^2 \Bigg\rangle - F_s^{2}(t_1, t_2) \Biggr]\,;
    \ee
    \item Finally, the non-Gaussian parameter $\alpha_2(t_1,t_2)$ (which is zero for Gaussian diffusion), 
    \be\label{eq:def_alpha2}
    \alpha_2(t_1, t_2) = \frac{3 \ea{(\Delta r)^{4}(t_1, t_2)}}{5 \Big(\ea{(\Delta r)^{2}(t_1, t_2)} \Big)^{2}}-1 \,.
    \ee
\end{enumerate}

\begin{figure}[H]
    \centering
       \includegraphics[width=0.495\textwidth]{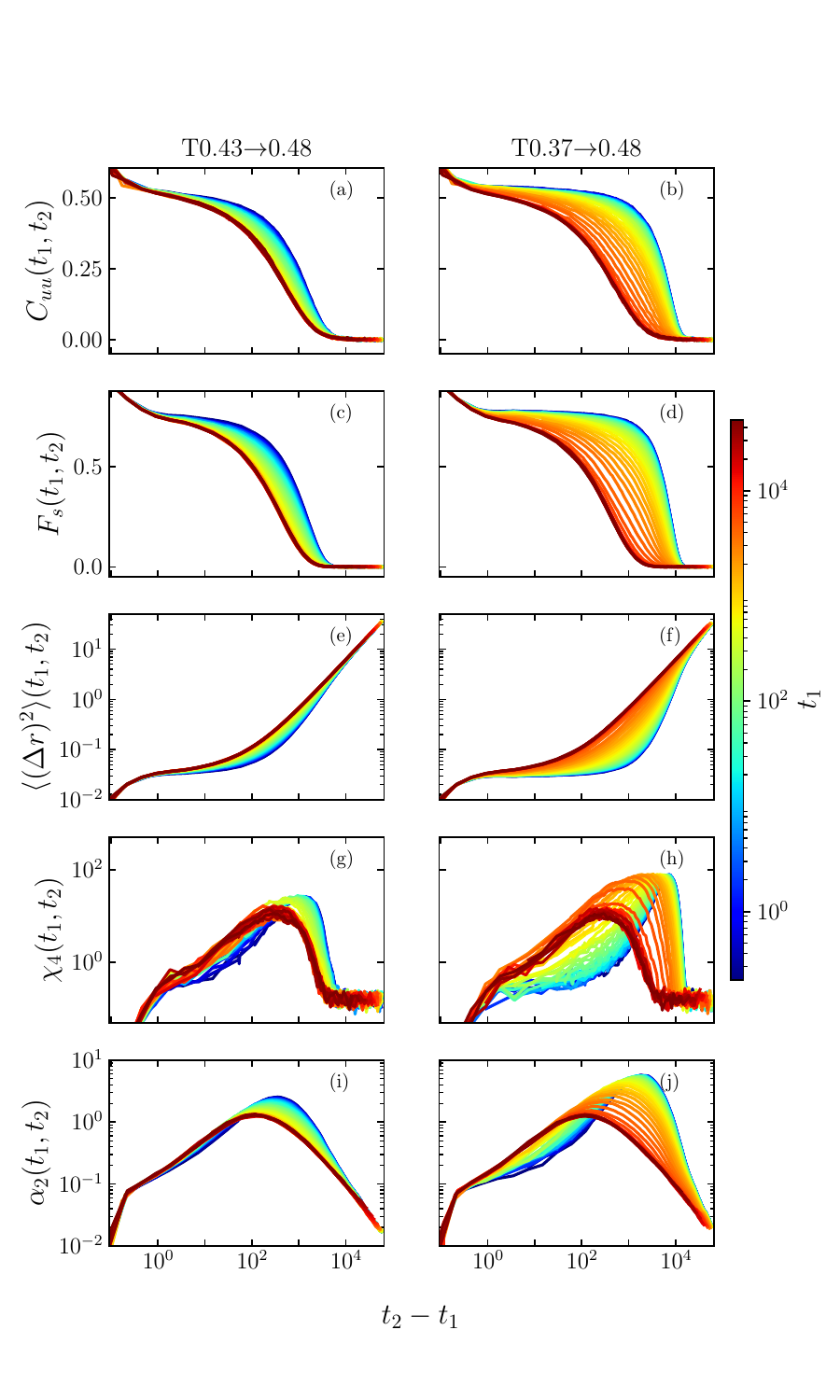}
       \includegraphics[width=0.495\textwidth]{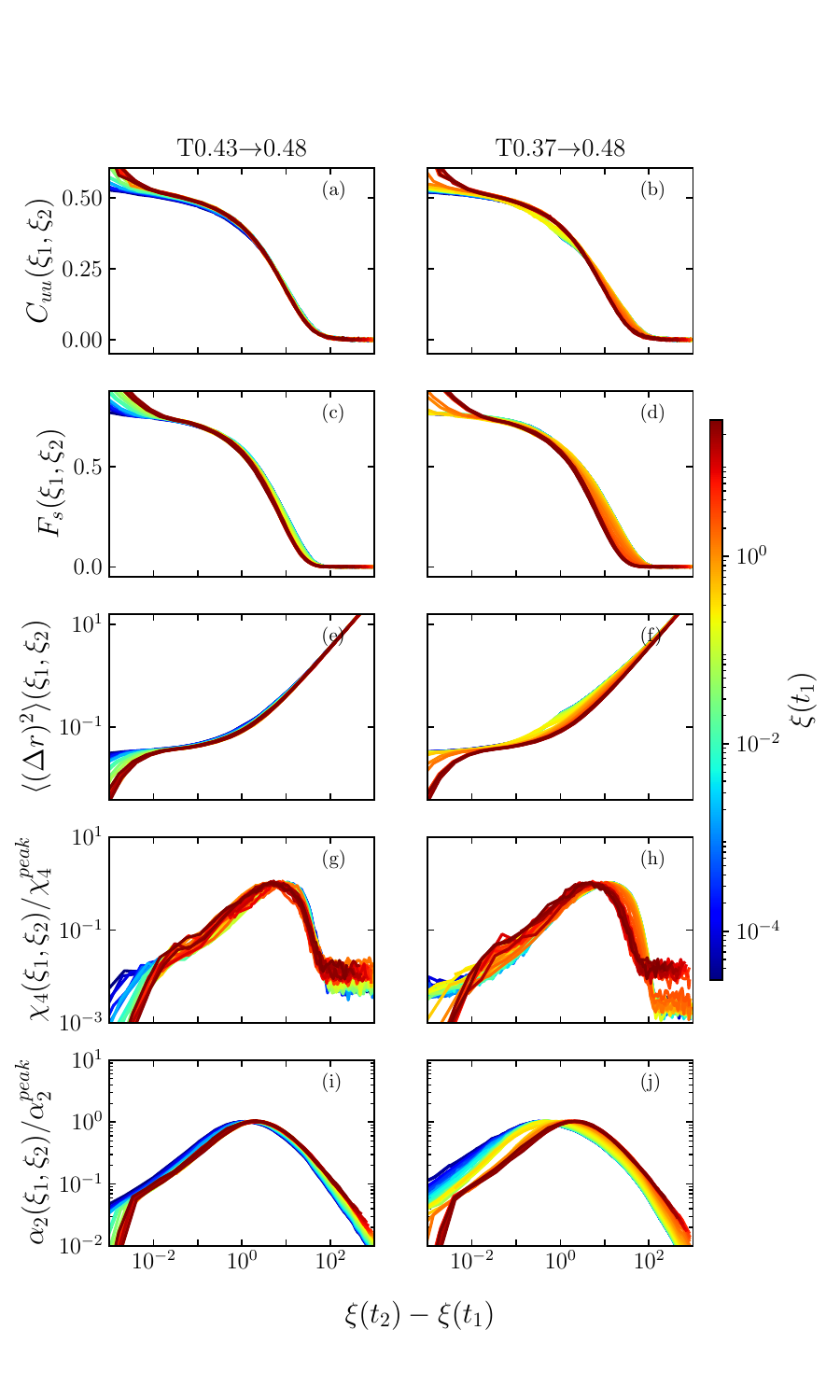}
       \caption{\textbf{Left} Two-time functions plotted during aging as functions of the time difference $t_2-t_1$. The left column gives data for the smaller up jump, the right for the larger one. The color of the curves vary from blue for small values of $t_1$ to red as this value increases. The following two-time observables are considered as functions of $t_2-t_1$:
       (a-b) Per-particle potential energy time-autocorrelation function;
       (c-d) Self part of the intermediate scattering function; 
       (e-f) Mean-square displacement;
       (g-h) Dynamical susceptibility;
       (i-j) Non-Gaussian parameter. 
       The curves are identical at large $t_1$ for both jumps because they share the same final temperature while they differ at small $t_1$.
       \textbf{Right} Same data plotted as functions of the material-time difference $\xi_2-\xi_1$, this time with peak normalisation of $\chi_4$ and $\alpha_2$.   
       There is a significantly better collapse for the smaller up jump than for the larger one.     
       }\label{fig:results}
\end{figure}

\Fig{fig:results} reports the results of our simulations for temperature up jumps from $T=0.43$ and $T=0.37$ to $T=0.48$. The left two columns show the time-autocorrelation functions during aging plotted as functions of $t_2-t_1$ with the smaller jump to the left, parametrized by the waiting time $t_1$. As expected, curve collapse is not observed for any of the five quantities. Instead a general speed up is seen, with the short-waiting time (bluish) curves moving to shorter times. This is the well-known ``self-catalyzed'' behavior, reflecting the fact that the structure gradually approaches that of $T=0.48$ at which relaxation is much faster than at the two starting temperatures. Comparing the results for the jump from $T=0.43$ to those of the $T=0.37$ to $T=0.48$ jump, there is a considerably larger spread in the latter case. This reflects that fact that the relaxation rates vary more, the larger the difference is between the initial and final temperatures of a jump. At the longest times corresponding to the red curves, a beginning collapse is observed because equilibrium at $T=0.48$ is approached.

The right two columns of \fig{fig:results} give the main results of this paper, investigating to which extent the material time defined from the potential-energy time-autocorrelation function collapses all the two times functions to be unique functions of $\xi_2-\xi_1$ in which $\xi_j$ is the material time at time $t_j$. Note that the non-monotonic quantities, the dynamical susceptibility $\chi_4$ and the non-Gaussian parameter $\alpha_2$, are normalised by their peak values; this is because we are only interested in the evolution of the shape of two-time correlations during aging. There is no collapse at short material times because, as already mentioned, the material-time concept makes little sense on the short time scale corresponding to phonon and slightly longer times. At larger values of $\xi_2-\xi_1$ there is a much better collapse, though to a varying degree and, curiously, the bluish curves corresponding to short waiting time collapse better than the curves corresponding to intermediate values of $t_1$. Generally, the smaller temperature jump results in a better collapse than the larger one. 

Starting with the smaller jump, all quantities show quite good collapse, apart from $\chi_4$ and $\alpha_2$ that do not collapse well (though much better so than in the real-time plot). The situation is worse for the large up jump in which case none of the five quantities exhibit a close to perfect collapse. Even here, however, most of the huge variation of the real-time panel is eliminated by the introduction of the material time. For the large jump, also the potential‑energy time-autocorrelation function itself displays small slope differences preventing a full collapse. This is interesting because according to Fig. 2(c), the triangular relation is obeyed quite well. Apparently, subtle deviations sum to a material-time clock not working well. This point deserves investigation in future works. -- One also observes that while the $\chi_4$ curves systematically shift toward later times compared to the final equilibrium value, the $\alpha_2$ curves behave oppositely: those at small $t_1$ relax faster than those at larger $t_1$. $\alpha_2$ probes the appearance of flow events that correspond to large deviations from the Gaussian dynamics, while $\chi_4$ reflects when those events become spacially correlated \cite{bir22, ton05}. One may thus speculate that the difference in behavior between $\chi_4$ and $\alpha_2$ might reflect heterogeneity occurring before cooperativity in the particles dynamics.

\section{Discussion}

We have shown that the triangular relation is well obeyed for the potential energy. Despite this, the material time defined from the potential energy is not able to describe well the aging of the five autocorrelation functions following very large temperature up jumps. Two possible explanations of this observation comes to mind. One possibility is that each quantity has its own material time in the sense that aging is better described in this way than by using a common material time. Another possibility is that dynamic heterogeneities invalidate the existence of any global material time for large up jumps, as suggested some time ago by Castillo and Parsaeian \cite{cas07}. This would make sense physically because little relaxation can take place in slow regions. Indeed, for vapour deposited ultra-stable glasses Vila-Costa \textit{et al.} have recently observed the existence of growing regions of activity during large temperature up jumps \cite{vil23, rui23}, and this observation has been reproduced in simulations \cite{her23}. On the other hand, even equilibrium highly viscous liquids exhibit dynamic heterogeneities, and here there is time-translational invariance of all time-autocorrelation functions and no need to introduce local times. Thus if local clocks are needed to describe large temperature up jumps, this can only be the case under specific aging conditions. -- In conclusion, more work is needed to clarify whether the introduction of local clocks can save the TN formalism or it must be discarded altogether for large temperature up jumps.

\begin{acknowledgments}
The authors thank Thomas Schr{\o}der for providing configurations equilibrated at T=0.37. 
This work was supported by the VILLUM Foundation's \textit{Matter} grant (No. VIL16515).
\end{acknowledgments}

\end{document}